\documentstyle[aps,epsf,multicol,prl]{revtex}
\begin{document}
\draft

\title{The Magnetic Susceptibility of Non-Interacting Nanoparticles}
\author{L. A. Ponomarenko, A. de Visser and E. Br\"{u}ck}
\address{Van der Waals-Zeeman Instituut, Universiteit van Amsterdam,
Valckenierstraat 65,
1018 XE Amsterdam,
The Netherlands}
\author{A. M. Tishin}
\address{
Physics Faculty, 
Moscow State University, Moscow, 119899, Russia}

\date{\today}
\maketitle
\begin{abstract}
We have calculated the low-field magnetic susceptibility $\chi$ of a system consisting of non-interacting mono-dispersed nanoparticles using a classical statistical approach. The model makes use of the assumption that the axes of symmetry of all nanoparticles are aligned and oriented at a certain angle $\psi$ with respect to the external magnetic field. An analytical expression for the temperature dependence of the susceptibility $\chi(T)$ above the blocking temperature is obtained. The derived expression is a generalization of the Curie law for the case of anisotropic magnetic particles. We show that the normalized susceptibility is a universal function of the ratio of the temperature over the anisotropy constant for each angle $\psi$. In the case that the easy-axis is perpendicular to the magnetic field the susceptibility has a maximum. The temperature of the maximum allows one to determine the anisotropy energy. 
\end{abstract}


\begin{multicols}{2}

\section{Introduction}

The theoretical investigation of magnetic single-domain particles goes back to the pioneering work of Stoner and Wohlfarth \cite{Stoner}. In the last decade, the interest in fine magnetic particles is growing, because of the prospect of application in high-density data storage devices (see for example Ref.~\onlinecite{Weller}). The size of the magnetic domains depends on the particular material and preparation method. However, particles with dimensions less than one micrometer, i.e. nanosized particles, usually consist of a single magnetic domain \cite{Mallinson}. Therefore, the term ``magnetic nanoparticles'' is normally used for fine single-domain particles.  

The remarkable properties of magnetic nanoparticles are due to their anisotropy. The anisotropy has its origin in the non-spherical shape of the particle (shape anisotropy) or in spin-orbit interaction. In the most simple model, the anisotropic nanoparticle can be represented by an ellipsoid of revolution. In this case, the anisotropy is described by one single parameter $K$, which is called the anisotropy constant. This constant $K$ can be calculated exactly from the ratio of the length of the minor and major axis of the ellipsoid \cite{Bozorth}. At the blocking temperature $T_b = KV/25k_B$ ($V$ is the volume of the particle and $k_B$ is Boltzmann's constant) the nanoparticle shows non-equilibrium properties on the time scale of the order of 100 seconds \cite{Weller}. In principle, magnetic particles are suitable for recording only at temperatures below $\sim T_b/2$. This is why a large number of theoretical and experimental papers have been devoted to the investigation of magnetic nanoparticles at temperatures below $T_b$. However, up to now only a few theoretical investigations have been reported focusing on the properties of nanoparticles at temperatures above $T_b$ \cite{Yasumori,Cregg1,Chantrell,Cregg2,Respaud}.

It is well known that the temperature dependence of the low-field susceptibility $\chi(T)$ of a system of non-interacting nanoparticles with randomly orientated axes of  symmetry follows the Curie law \cite{Yasumori}. In this case, $\chi(T)$ does not contain any information about the anisotropy of the nanoparticles. However, the time-averaged magnetization of an isolated particle is obviously anisotropic. The susceptibility of a system of non-interacting nanoparticles with parallel axes of symmetry was analytically calculated only for  several special cases \cite{Cregg1}. The purpose of the present paper is to report on the temperature dependence of the low-field susceptibility in the general case of an arbitrary angle between the symmetry axis and the external magnetic field.

\section{Model}

Let us consider a system consisting of a large number of identical non-interacting nanoparticles with magnetic moment $\bbox{\mu}$ and volume $V$ and all axes of symmetry parallel to each other. Since the interactions are neglected this many-particle problem can be reduced to that of a single particle. We assume that the particle has the shape of an ellipsoid of revolution with an anisotropy constant $K$. The spin-orbit contribution to the anisotropy is assumed to be zero. For a fixed orientation of the particle in space, the axis of symmetry makes an angle $\psi$ with respect to the external magnetic field $\bbox{H}$. The potential energy of the particle is then given by
\begin{equation}
E = - KVcos^2 \theta - \mu H cos\omega, \label{energy}
\end{equation}
where $\theta$ is the angle between the magnetic moment and the applied field and $\omega$ is the angle between the axis of symmetry and the applied field (see Fig.1). The first term in Eq.~(\ref{energy}) yields the anisotropy energy, while the second term represents the Zeeman energy of the magnetic dipole in the external field. Our assumption of non-interacting particles is valid if the energy of the dipole-dipole interaction 
$\bbox{\mu}_i \bbox{\mu}_j/r^3_{ij}$ ($\bbox{\mu}_i$ and $\bbox{\mu}_j$ are magnetic moments 
of two neighboring particles and $r_{ij}$ is the interparticle distance) is much smaller than all other energies occurring in the problem, namely the anisotropy energy $KV$, the Zeeman energy $\mu H$ and the thermal energy $k_B T$. 

\begin{figure}[b]
\begin{center}
\setlength{\unitlength}{1mm}
\begin{picture}(80,70)(0,0)
\put(0,0)
{\epsfxsize=7cm{\epsffile{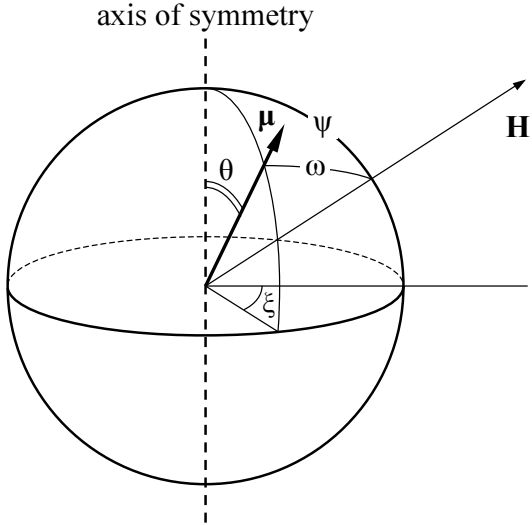}}}
\end{picture}
\vspace{5 mm}
\caption{
Spherical coordinate system used to calculate the susceptibility of a single magnetic nanosized particle. $\theta$ is the angle between the magnetic moment $\bbox{\mu}$ and the axis of symmetry (dashed vertical line), $\omega$ is the angle between the magnetic moment and the field direction $\bbox{H}$, $\psi$ is the angle between $\bbox{H}$ and the axis of symmetry, and $\xi$ defines the angle between the projections of $\bbox{\mu}$ and $\bbox{H}$ on the equatorial plane. 
}
\label{Fig. 1}
\end{center}
\end{figure}

For $K$ positive the energetically favorable direction of magnetization (at zero field) is parallel to the symmetry axis. Hence, the case $K > 0$ corresponds to the``easy-axis" type of particle, while the case $K < 0$ represents the``easy-plane" type of particle. The difference between these two cases is shown in Fig.~2. The prolate shape of the particle results in a positive anisotropy constant. On the contrary, the oblate shape gives rise to a negative value of the anisotropy constant $K$. The theory proposed here enables one to describe both cases, however, we concentrate mainly on the``easy-axis" type of particles. Since``easy-plane" particles do not show blocking of the magnetic moment at any temperature, these particles cannot be used for recording. The magnetization of ``easy-plane" nanoparticles will be briefly discussed in a separate subsection.

\begin{figure}
\begin{center}
\setlength{\unitlength}{1mm}
\begin{picture}(80,70)(0,0)
\put(0,0)
{\epsfxsize=7cm{\epsffile{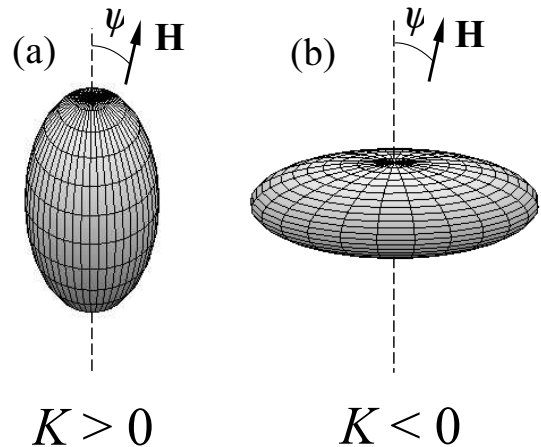}}}
\end{picture}
\vspace{7 mm}
\caption{
Schematic representation of magnetic nanoparticles with opposite signs of anisotropy constant. The anisotropy is due to the ellipsoidal shape of the particle. The prolate shape ({\it a}) gives rise to a positive value of the anisotropy constant $K$. The oblate shape ({\it b}) gives rise to a negative $K$. $\psi$ is the angle between the axis of 
symmetry and the external magnetic field $\bbox{H}$.
}
\label{Fig. 2}
\end{center}
\end{figure}

Above the blocking temperature the magnetic nanoparticle is in an equilibrium state, and, therefore, we can use a classical statistical approach and classical thermodynamics to calculate the magnetic properties. Thus our results are of relevance above $T_b$. 

By introducing the parameters $\alpha = KV / (k_BT)$ and $\beta = \mu H / (k_BT)$, the partition function for a single domain particle can be written as \cite{Yasumori}:
\begin{equation}
Z = {1 \over 4\pi} \int^{2\pi}_{0} \int^{\pi}_{0} \exp(\alpha \cos^2\theta+\beta \cos\omega)\sin\theta d\theta d\xi. \label{Z1}
\end{equation}

The angle $\omega$ can be expressed through $\psi$ and the integration variables $\theta$ and $\xi$ by the relation:
\begin{equation}
\cos \omega = \cos \theta \cos \psi + \sin \theta \sin \psi \cos \xi. \label{omega}
\end{equation}

Once the analytical expression for the partition function $Z(\alpha, \beta, \psi)$ is known one can  calculate the component of the equilibrium magnetization $M_{\|}$ parallel to the field \cite{Chantrell}:
\begin{equation}
M_{\|} = \mu {1 \over Z} {\partial Z \over \partial \beta}. \label{magnetization}
\end{equation}

In order to obtain the analytical expression for $Z(\alpha, \beta, \psi)$, the double integral in Eq.~(\ref{Z1}) should be calculated first. By reducing the double integral for the partition function in a single integral \cite{Cregg2} it follows:
\begin{eqnarray}
Z = \int^{\pi/2}_{0} \exp(\alpha \cos^2 \theta) \cosh(\beta \cos\theta\cos\psi) \nonumber\\
\times I_0(\beta\sin\theta\sin\psi)sin\theta d\theta, \label{Z2}
\end{eqnarray}
where $I_0(\cdot)$ is the modified zero-order Bessel function of the first kind. The same expression was also derived as an intermediate result in Ref.~\onlinecite{Garcia}. We used Eq.~(\ref{Z2}) as the starting point for our calculations. 

\section{The partition function}

By inserting the expansions for $I_0(\cdot)$ and $\cosh(\cdot)$ into Eq.~(\ref{Z2}), and by switching the order of integration and summation, we obtain the partition function as a infinite series (see Appendix A):
\begin{equation}
Z(\alpha, \beta, \psi) = \sum^{\infty}_{n=0}{(\beta/2)^{2n} \over n!\Gamma(n+3/2)} \Omega_n(\alpha, \psi), \label{Z3}
\end{equation}
where $\Omega_n(\alpha, \psi)$ represents the internal sum:
\begin{eqnarray}
\Omega_n(\alpha, \psi) = \sum^{n}_{k=0}{n! \over k!(n-k)!} \cos^{2k}\psi \cdot \sin^{2(n-k)} \psi \nonumber \\
\times M(k+{1 \over 2}, n + {3 \over 2}, \alpha). \label{Omega}
\end{eqnarray}
Here $M(\cdot,\cdot,\cdot)$ is the Kummer function (confluent hypergeometric function) and $\Gamma(\cdot)$ is the gamma function. The expressions (\ref{Z3}) and (\ref{Omega}) for the partition function are similar to the ones obtained in Ref.\onlinecite{Garcia}, where the non-linear magnetisation of superparamagnetic particles with a random orientation of easy axes is analysed. Notice, that the function $\Omega_n(\alpha, \psi)$ does not contain the parameter $\beta$. This significantly simplifies the calculation of the derivative $\partial Z/ \partial \beta$:
\begin{equation}
{\partial Z \over \partial \beta} = \sum^{\infty}_{n=0} {(\beta/2)^{2n-1} \over (n-1)! \Gamma (n+3/2)} \Omega_n (\alpha, \psi), \label{dZ}
\end{equation}

Both series in the expressions for $Z$ and $\partial Z/ \partial \beta$ converge at all values of $\alpha$, $\beta$ and $\psi$. The reminder of the $N$-th partial sum for the partition function and its derivative, denoted by $R_{N+1}\{Z\}$ and $R_{N+1}\{\partial Z/\partial \beta\}$, respectively, can be estimated as:
\begin{equation}
R_{N+1} \{ Z \} < {(\beta/2)^{2N} \over [(N+1)!]^2} \exp \biglb( | \alpha | + {\beta \over 2} \bigrb ), \label{RZ}
\end{equation}

\begin{equation}
R_{N+1} \Bigl\{ {\partial Z \over \partial \beta} \Bigr\} <  {( \beta /2)^{2N+1} \over (N!)^2} \exp \biglb( | \alpha | + {\beta \over 2} \bigrb ). \label{RdZ}
\end{equation}

These two inequalities are important for estimating the errors in the numerical calculations.

\section{Low field susceptibility}

In the low-field case $\mu H \ll k_B T$ ($\beta \ll 1$), only the leading terms in (\ref{Z3}) and (\ref{dZ}) are retained. This simplification makes it possible to derive an analytical expression for the low-field susceptibility (see Appendix B):

\begin{figure}
\begin{center}
\setlength{\unitlength}{1mm}
\begin{picture}(80,70)(0,0)
\put(0,0)
{\epsfxsize=7cm{\epsffile{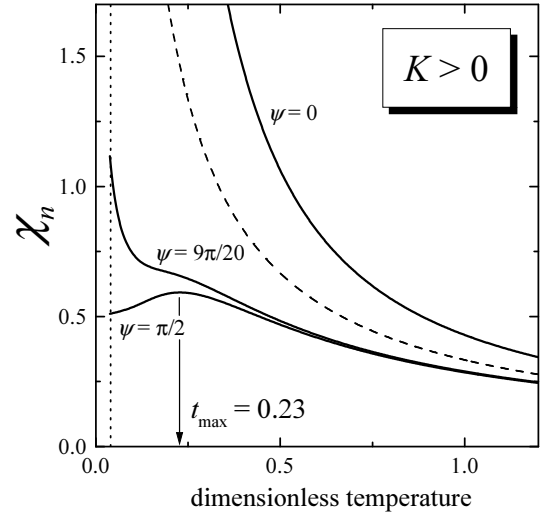}}}
\end{picture}
\vspace{5 mm}
\caption{
The normalized susceptibility $\chi_n = \chi \cdot KV / \mu^2$ as a function of the dimensionless temperature $t = k_B T/(KV)$ for $K > 0$ at several values of the angle $\psi$ between the easy axis and the direction of the magnetic field, as indicated. The vertical dotted line at $t_b = 1/25$ represents the blocking temperature, below which the presented model is no longer valid.  The dashed line corresponds to the Curie law.
}
\label{Fig. 3}
\end{center}
\end{figure}

\begin{equation}
\chi = {\mu^2 \over KV} \cdot \alpha \biggl\{ \cos^2 \psi + \Bigl({1 \over 3} - \cos^2 \psi\Bigr) {M({1 \over 2}, {5 \over 2}, \alpha) \over M({1 \over 2}, {3 \over 2}, \alpha)} \biggr\}. \label{chi_final}
\end{equation}

Notice, that Eq.~(\ref{chi_final}) represents the component of the magnetic susceptibility parallel to the magnetic field per particle. To obtain the volume susceptibility the last equation should be multiplied by the density of particles. Eq.~(\ref{chi_final}) is a generalization of the Curie law for the case of an anisotropic magnetic particle. We stress that the function to calculate the magnetic susceptibility derived here is an analytical expression. Although the Kummer function $M(\cdot,\cdot,\cdot)$ is not frequently used, its properties are well known. The confluent hypergeometric functions are available in most mathematical software packages. The validity of Eq.~(\ref{chi_final}) can easily be checked for two special cases: zero anisotropy ($K = 0$) and randomly orientated symmetry axes. In both cases, it gives the Curie law expression $\chi = \mu^2/(3 k_B T)$. This is in agreement with the results presented in Ref.\onlinecite{Yasumori}.

In order to compare the calculated susceptibility with the result obtained by Chantrell {\it et al.} \cite{Chantrell}, we use the integral representation of the Kummer function \cite{Abramovitz} and transform Eq.~(\ref{chi_final}) into
\begin{equation}
\chi = {\mu^2 \over k_B T} \Bigl\{ \cos^2 \psi + {1 \over 2}(1 - 3\cos^2 \psi) \Bigl(1+ {1 \over 2\alpha} - {e^\alpha \over \alpha I(\alpha)}\Bigr) \Bigr\}, \label{chi_chantrell}
\end{equation}
where $I(\alpha) = \int^1_{-1} \exp(\alpha x^2) dx$. Our Eq.~(\ref{chi_chantrell}) differs from the result given by Chantrell {\it et al.} in the sign of the $1/(2 \alpha)$ term (see Eq.~(7) in Ref.~\onlinecite{Chantrell}). Notice, the temperature dependence of the susceptibility was not examined in that paper.

\section{Discussion and analysis}

\subsection{Case of positive $K$}

Let us introduce the new dimensionless quantity 
\begin{equation}
\chi_n =  \chi {K V \over \mu^2}, \label{chi_norm}
\end{equation}
which we call the normalized susceptibility. According to Eq.~(\ref{chi_final}), $\chi_n$ is a function of the two parameters $\alpha$  and $\psi$. The anisotropy constant $K$ and the temperature $T$ enter in the expression for $\chi_n$ only as a quotient through the parameter $\alpha = KV/(k_B T)$. Therefore, the general behavior of the temperature dependence of the normalized susceptibility is the same for different nanoparticles with different anisotropy constants, but with the same value of the angle $\psi$. The effect of the constant $K$ is to change the temperature scale only. Therefore, we show in Fig.~3 the normalized susceptibility $\chi_n$ as a function of the dimensionless temperature $t = |k_B T / (K V)| = |\alpha^{-1}|$ calculated for the case $K > 0$ (``easy axis'') at several values of the angle $\psi$. For comparison we also show the susceptibility for particles with randomly oriented directions of easy axes. As we mentioned above, the latter susceptibility obeys the Curie law. From Fig.~3 it follows that the low-field susceptibility for the parallel ($\psi = 0$) and perpendicular ($\psi = \pi/2$) orientations of the easy axis with respect to the external field show very different behavior. For $\psi = 0$ the susceptibility values exceed the ones obtained by the Curie law at all temperatures. On the contrary, for $\psi = \pi/2$ $\chi_n(t)$ is always smaller than obtained by the Curie law. This qualitative difference is not very surprising and can be explained in a simple picture as follows. If the field direction coincides with the easy axis ($\psi = 0$) the anisotropy forces the magnetic moment to align along the external field. As a result the averaged component of the magnetic moment projected on the field direction is larger. In the perpendicular configuration, i.e. the external field perpendicular to the easy axis ($\psi = \pi/2$), the magnetic moment of the particle is deflected from the direction of the magnetic field by the anisotropy, which results in a smaller averaged magnetization.

A quite important result is that $\chi_n(t)$ goes through a maximum at $t_{max} = 0.23$ in the case $\psi = \pi/2$. The value of $t_{max}$ corresponds to the temperature:
\begin{equation}
T_{max} = 0.23KV/k_B. \label{Tmax}
\end{equation}

Using Eq.~(\ref{Tmax}), the anisotropy energy $KV$ can be determined in a simple way from the experimental curve of the temperature dependence of the low-field susceptibility. However, the maximum becomes weaker and finally disappears, when the angle $\psi$ deviates from $\pi/2$, i.e. when the easy axis is not exactly perpendicular to the field. For example, for $\psi = 9 \pi/20$ this deviation is only $9^\circ$, but $\chi_n(t)$ no longer exhibits a maximum, as shown in Fig.~3.

\begin{figure}
\begin{center}
\setlength{\unitlength}{1mm}
\begin{picture}(80,70)(0,0)
\put(0,0)
{\epsfxsize=7cm{\epsffile{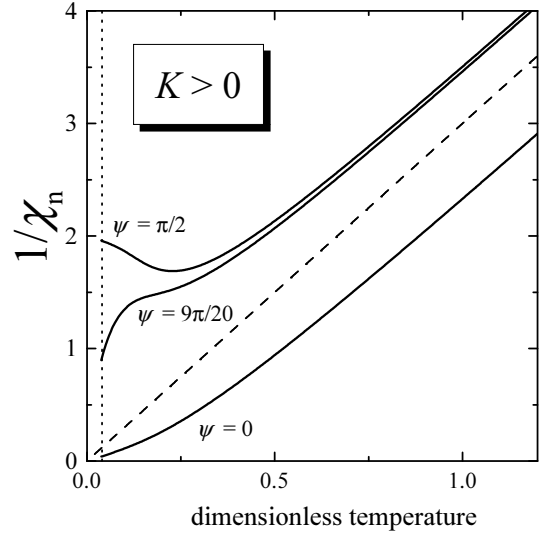}}}
\end{picture}
\vspace{5 mm}
\caption{
The~normalized~reciprocal~susceptibility
$\chi_n^{-1} = \chi^{-1} \mu^2/ (KV)$ as a function of the dimensionless temperature $t = k_B T/(KV)$ for $K > 0$ at several values of the angle $\psi$ between the easy axis and the direction of the magnetic field, as indicated. The dashed line corresponds to the Curie law.
}
\label{Fig. 4}
\end{center}
\end{figure}

The temperature dependence of the reciprocal dimensionless susceptibility $\chi_n^{-1}(t)$ is shown in Fig.~4, for the same values of the angle $\psi$ as used in Fig.~3. In the high temperature region the reciprocal susceptibility follows a Curie-Weiss law $\chi_n^{-1} = 3(t-\theta_d)$. The dimensionless Weiss constant $\theta_d$ is a function of the angle $\psi$. The sign of $\theta_d$ is positive for $\psi = 0$ and negative for $\psi = \pi/2$. The Curie-Weiss law and the angular dependence $\theta_d(\psi)$ can be evaluated from Eq.~(\ref{chi_final}). The case of high temperatures corresponds to small values of the parameter $\alpha$. Using an expansion of the Kummer function (see Eq.~(13.1.2) in Ref.\onlinecite{Abramovitz}) the reciprocal susceptibility $\chi_n^{-1}$ for the case $\alpha \ll 1$ can be written as
\begin{equation}
\chi_n^{-1} =  {3 \over \alpha} + {2 \over 5} (1 - 3\cos^2 \psi), \label{CurieWeiss}
\end{equation}
from which it immediately follows 
\begin{equation}
\theta_d = - {2 \over 15} (1 - 3\cos^2 \psi). \label{thetad}
\end{equation}

\begin{figure}
\begin{center}
\setlength{\unitlength}{1mm}
\begin{picture}(80,70)(0,0)
\put(0,0)
{\epsfxsize=7cm{\epsffile{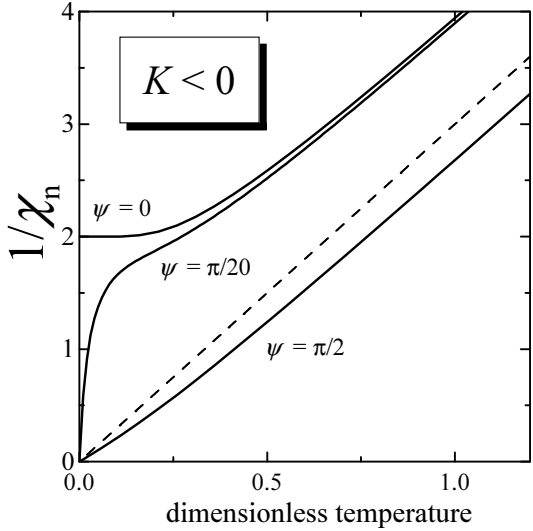}}}
\end{picture}
\vspace{5 mm}
\caption{
The~normalized~reciprocal~susceptibility 
$\chi_n^{-1} = \chi^{-1} \mu^2/ (KV)$ as a function of the dimensionless temperature $t = k_B T/(KV)$ for $K < 0$ at several values of the angle $\psi$, as indicated. The dashed line corresponds to the Curie law.
}
\label{Fig. 5}
\end{center}
\end{figure}

\subsection{Case of negative K}

The model proposed here can also be applied to calculate the low-field susceptibility of ``easy-plane'' type particles. In Fig.~5 we show the calculated temperature variation of 
$\chi_n^{-1}(t)$ for several values of $\psi$. Unlike the case of $K > 0$, the magnetic susceptibility of an``easy-plane''-type particle is a monotonic function of temperature for all $\psi$. Such a particle does not exhibit blocking of the magnetic moment, because states with minimal potential energy are not separated by an anisotropy barrier at zero field. 

\section{Concluding remarks}

In this paper we have presented a simple model to calculate the susceptibility of nanosized particles above the blocking temperature. The model describes the case of monodispersed nanoparticles with a perfect alignment of the easy axes. We have evaluated the susceptibility for arbitrary values of the angle $\psi$, i.e. the angle between the field direction and the easy axis. One of the main results of our calculations is that the normalized susceptibility $\chi_n$ {\it versus} the dimensionless temperature $t$ is a universal function for each $\psi$, which does not depend on the particular value of the anisotropy constant $K$. One of the advantages of the model is that only one phenomenological parameter is involved, namely the anisotropy energy $KV$. 

In practice, samples consist of nanoparticles with a significant volume distribution. Moreover, the easy axes are usually randomly orientated. As we mentioned above, randomly orientated easy axes impede the observation of anisotropy at temperatures larger than the blocking temperature. A significant volume distribution casts serious doubts on the applicability of simple models used to examine the measured susceptibility. We are quite confident, however, that the current efforts in improvement of the sample preparation process of nanoparticles with a narrow size distribution, as well as the development of a technique for magnetic orientation of nanoparticles \cite{Bentivegna}, will enable experimental tests of the model proposed here in the near future.

\acknowledgements

The authors have benefited from stimulating discussion with M. G. Mikheev, D. T. N. de Lang, S. P. Gubin and A. K. Zvezdin. We are grateful to A. S. Mischenko for numerical calculations which he made to check some of the presented results. The work was partly supported by RFBR grant 01-02-17703.

\appendix

\section {}

Using expansions for $I_0(\cdot)$ and $\cosh(\cdot)$ (see Eq.~(9.6.12) and Eq.~(4.5.63) from Ref.~\onlinecite{Abramovitz}) the partition function can be written as:
\begin{eqnarray}
Z(\alpha, \beta, \psi) = \int^{\pi/2}_{0} \exp (\alpha \cos^{2} \theta) \sum^{\infty}_{i=0}{(\beta \cos \theta \cos \psi)^{2i} \over (2i)!} \nonumber \\
\times \sum^{\infty}_{j=0}{({1 \over 2} \beta \sin \theta \sin \psi)^{2j} \over (j!)^2} 
\sin \theta d\theta. \label{app1}
\end{eqnarray}
By changing the order of summation and integration and using relation $(2i)! = 2^{2i}i! \Gamma (i+1/2)/\sqrt{\pi}$ , which can be obtained from Eq.~(6.1.12) in Ref.~\onlinecite{Abramovitz} we arrive at:
\begin{eqnarray}
Z(\alpha, \beta, \psi) = \sqrt{\pi} \sum^{\infty}_{i=0} 
{(\beta \cos \psi)^{2i} \over 2^{2i} i! \Gamma(i + {1 \over 2})} 
\sum^{\infty}_{j=0} {({1 \over 2} \beta \sin \psi )^{2j} \over (j!)^2} \nonumber \\
\times \int^{\pi/2}_{0} \exp (\alpha \cos^{2} \theta) \cos^{2i} \theta \sin^{2j+1} \theta d\theta. \label{app2}
\end{eqnarray}
The integral in Eq.~(\ref{app2}) can be transformed to the integral representation of Kummer function (see Eq.~(13.2.1) in Ref.~\onlinecite{Abramovitz}) by substituting $t = \cos^2 \theta$. Then the partition function is:
\begin{eqnarray}
Z(\alpha, \beta, \psi) = \sqrt{\pi} 
\sum^{\infty}_{i=0} {({1 \over 2} \beta \cos \psi)^{2i} \over i!} 
\sum^{\infty}_{j=0} {({1 \over 2} \beta \sin \psi )^{2j} \over j!} \nonumber \\
\times {M(i+{1 \over 2}, i+j+{3 \over 2}, \alpha) \over \Gamma(i+j+{3 \over 2})}. \label{app4}
\end{eqnarray}
By changing the order of summation ($k = i$, $n = i +j$) to gather terms with the same power of the parameter $\beta$ we obtain:
\begin{eqnarray}
Z(\alpha, \beta, \psi) = \sqrt{\pi} 
\sum^{\infty}_{n=0} \sum^{n}_{k=0}{({1 \over 2} \beta \cos \psi)^{2k} \over k!} 
{({1 \over 2} \beta \sin \psi )^{2(n-k)} \over (n-k)!} \nonumber \\
\times {M(k+{1 \over 2}, n+{3 \over 2}, \alpha) \over \Gamma(n+{3 \over 2})}. \label{app5}
\end{eqnarray}
Last equation can also be written as:
\begin{eqnarray}
Z(\alpha, \beta, \psi) = \sqrt{\pi} 
\sum^{\infty}_{n=0} {({1 \over 2} \beta)^{2n} \over n! \Gamma(n+{3 \over 2})}
\sum^{n}_{k=0}{n! \over k!(n-k)!} \nonumber \\
\times cos^{2k} \psi \sin^{2(n-k)} \psi
\cdot M(k+{1 \over 2}, n+{3 \over 2}, \alpha). 
\label{app6}
\end{eqnarray}
Since multiplying the partition function by any constant does not affect the magnetization the prefactor $\sqrt{\pi}$ can be omitted. Eq.~(\ref{app6}) is identical to the expression for $Z(\alpha, \beta, \psi)$ in Eq.~(\ref{Z3}), except for the prefactor $\sqrt{\pi}$.

\section {}

Considering only the leading terms in (\ref{Z3}) and (\ref{dZ}) we write:

\begin{equation}
Z(\alpha, \beta, \psi) = {\Omega_0(\alpha, \psi) \over \Gamma(3/2)}, \label{Zlow}
\end{equation}

\begin{equation}
{\partial Z \over \partial \beta} = {\beta \over 2} \cdot {\Omega_1(\alpha, \psi) \over \Gamma(5/2)}. \label{dZlow}
\end{equation}

The parallel component of the magnetization in low magnetic field is then given by:
\begin{equation}
M_{\|} = \mu {\beta \over 3} {\Omega_1(\alpha, \psi) \over \Omega_0(\alpha, \psi)}. \label{Mlow1}
\end{equation}

By substituting the expressions for $\Omega_0$ and $\Omega_1$, obtained from Eq.~(\ref{Omega}), and taking into account that $M({3 \over 2}, {5 \over 2}, \alpha) = 3 M({1 \over 2}, {3 \over 2}, \alpha) - 2 M({1 \over 2}, {5 \over 2}, \alpha)$ (see (13.4.3) in Ref.~\onlinecite{Abramovitz}), we arrive at the following expression for $M_{\|}$:
\begin{equation}
M_{\|} = \mu \beta \biggl\{ \cos^2 \psi + \Bigl({1 \over 3} - \cos^2 \psi\Bigr) {M({1 \over 2}, {5 \over 2}, \alpha) \over M({1 \over 2}, {3 \over 2}, \alpha)} \biggr\}. \label{Mlow2}
\end{equation}
From this last expression for $M_{\|}$ we obtain Eq.~(\ref{chi_final})  using the relationships $\beta = \alpha \mu H/ (K V)$ and $\chi = M_{\|}/H$.


\end{multicols}
\end{document}